# Preferential Composition during Nucleation and Growth in Multi-Principal Elements Alloys

Saswat Mishra and Alejandro Strachan

School of Materials Engineering and Birck Nanotechnology Center,

Purdue University, West Lafayette, Indiana 47907, USA

**Abstract**

The crystallization of complex, concentrated alloys can result in atomic-level short-range order, composition gradients, and phase separation. These features govern the properties of the resulting alloy. While nucleation and growth in single-element metals are well understood, several open questions remain regarding the crystallization of multi-principal component alloys. We use MD to model the crystallization of a five-element, equiatomic alloy modeled after CoCrCuFeNi upon cooling from the melt. Stochastic, homogeneous nucleation results in nuclei with a biased composition distribution, rich in Fe and Co. This deviation from the random sampling of the overall composition is driven by the internal energy and affects nuclei of a wide range of sizes, from tens of atoms all the way to super-critical sizes. This results in short range order and compositional gradients at nanometer scales.

## 1. Introduction

High entropy alloys (HEAs) and related materials [1] are a new class of metals with a range of attractive properties. These include high-temperature strength surpassing current superalloys [2] and improved radiation hardening. [3, 4] These materials are being studied for various applications. For example, CaMgZnSrYb alloy has been used to promote bone formation, and AlFeCoCrNiCu, CrFeNiCuMoCo, and AlCoCrFeNiTi have been used for applications that require good corrosion resistance. In addition, NbMoTaV has been used for high-strength applications in extreme conditions such as high temperatures and high radiation environments. [2] As with all materials, properties can be significantly tuned via microstructure engineering, either during crystallization or via subsequent heat treatments, and even short-range order can affect properties. [5] A fundamental material process at the heart of nearly all fabrication processes used for HEAs is crystallization from the melt. Homogeneous crystallization is governed by stochastic nucleation and grain growth. These processes determine the microstructure of the resulting material, which, in turn, plays a key role in its properties and can affect annealing treatments. Despite the significant interest in these alloys, several questions remain open regarding crystallization, and in this paper, we use multi-million atom molecular dynamics (MD) simulations to characterize the underlying processes at the atomic scale.

During nucleation, thermal fluctuations in the liquid result in local crystalline order [6]. Most of these nuclei do not prosper, even at temperatures below the melting temperature, since small crystallites are energetically unfavorable. A temperature-dependent, critical nuclei size must be reached before crystal growth results in lower free energy. Both nucleation and growth are affected by temperature (and consequently cooling rate) and the presence of impurities that can act as nucleating sites, decreasing the energy barrier. Nucleation and grain growth in HEAs involve an additional complication compared to traditional alloys or single-element metals: their multi-principal component nature enables a wide range of possible atomic arrangements and compositions for the nuclei. These fluctuations of local chemistry can affect stability and nucleation barriers, and a local composition differing from that of the homogeneous liquid can affect the nucleation rate.

There have been significant advances in the understanding of nucleation since the development of classical nucleation theory that emerged from the works of Gibbs in the nineteenth century. [7] Despite experimental challenges associated with the time and length scales involved [8], metastable nucleation states have been observed [9,10] and this led to a better understanding of the dynamical nature of the nucleation process. For example, the nucleation process for magnetite involves a multistep pathway, including metastable disordered iron oxide formation, with no single route leading to the final crystallite. [11] Transient phases can help explain the precipitation behavior and morphology of the stable phase during crystallization but are difficult to detect. Time-resolved synchrotron x-ray scattering was used to observe a transient metastable phase during the triclinic phase crystallization of hexadecane. [12] Non-photochemical laser-induced nucleation also provided experimental verification for multistep nucleation. [13] Despite this progress, it remains difficult to distinguish between nucleation and growth in these experimental studies. [14] The study of such dynamical processes is further complicated in the case of HEAs. Hence, atomistic studies have played an important role in the study of crystallization [15-19] and phase transformation [20-24] in these materials. Studies of solid-solid transformations between BCC and the B2 phases [25-28] identified nucleation site preference of Hf-Zr and Nb-Ta pairs in HfNbTaTiZr using a MEAM potential. [29] Guo et al. [15] characterized the effect of local chemical order on solidification nucleation from the melt in AlCoCuFeNi using MD and found the Ni and Co regions to be easier to nucleate. Wang et al. studied a CoCrCuFeNi system using MD and observed high Fe concentration in the nucleus. [18]. However, the mechanisms and driving force behind this non-stoichiometric nucleation remain unknown, and few atomistic studies have been carried out on the nucleation behavior in their respective systems because of a lack of good interatomic potentials to enable large enough simulations. [28] This paper addresses these gaps.

## 2. Simulation details

The initial structure for our molecular dynamics simulations was obtained by creating an 80x80x80 supercell (2,048,000 atoms) with a face-centered cubic (FCC) unit cell. Atoms are assigned to lattice sites stochastically with a composition of 20% of each of the elements to create an equiatomic CoCrCuFeNi system. Starting from the initial crystal, a liquid structure is equilibrated at 3000 K and 1 atm using isobaric, isothermal MD for 100 ps. To simulate crystallization, we first quenched the system from 3000 K to 2000 K at 5 K/ps, thermalized the system at 2000K for another 100 ps, and then cooled it from 2000K at a rate of 5K/ps to the crystallization temperature of interest (1400 K,1350 K, and 1300 K). The system is then maintained at the desired temperature for 0.5 ns, enough to observe full crystallization. We used 5 independent and uncorrelated structures for each crystallization temperature to collect statistics. These independent structures were obtained by taking snapshots from the thermalized liquid simulation at 3,000K at different times (after 100 ps), separated by at least 10 ps to ensure that they were uncorrelated.

All simulations were performed with the Large-scale Atomic/Molecular Massively Parallel Simulator (LAMMPS) [30] with a timestep of 1 femtosecond. We used the Nosé-Hoover thermostat and Hoover barostat with relaxation times of 0.1 ps and 1 ps, respectively. Atomic interactions were described using an embedded atom model developed by Farkas et al. [31].

We used the average energy of each of the elements of the supercooled liquid (at t=0) to obtain the reference energy for the elements. We then used this reference energy to calculate the formation energy of the nuclei by summing over the atomic energy for each atom in the nuclei with reference to the average energies of the elements. We then divided the formation energy by the total number of atoms in the nuclei to obtain a formation energy per atom, which can then be compared with nuclei of different sizes.

## 3. Structure evolution during nucleation and growth

Figure 1 shows the resulting microstructure of an alloy crystallized at 1400 K, atoms are colored by their x-orientation in the quaternion representation. [32] We observe multiple grains within the simulation cell and a rich sub-grain structure. To understand the process of nucleation and growth that results in the observed microstructures, we analyzed the local structure around each atom and performed a cluster analysis to group neighboring crystalline atoms into crystallites or nuclei. The local atomic structure is characterized using the polyhedral template matching (PTM) analysis and classifies atoms into FCC, HCP, BCC, or amorphous environments. Figure 2 shows the time evolution of the fraction of atoms in each local structure, with the inset focusing on crystalline atoms. Crystallization is observed within the 500 ps duration of the simulation, and the analysis indicates a larger fraction of HCP atoms as compared to the stable FCC structure during the early, nucleation, stages of the process. We observe a small amount of BCC local structures. This suggests that the nucleation process is favored around HCP structures even though FCC is energetically favored in the bulk. As noted in the introduction, experimentally, it has been observed that HCP can be stabilized at smaller sizes, but it converts to FCC with increasing size, and our results align with this observation. Results for T=1350 K and 1300 K are included in the SI, where we observe the same HCP to FCC and BCC to FCC transformation. Some computational studies about homogeneous nucleation in HEAs exist in the literature. Zadeh et al. found that the FCC and BCC structures compete in CoFeNiPd. [33] Bahramyan et al. used Al in a CrCoFeCuNi system and reported BCC solidification only. [34] Gao et al. compared the homogeneous and heterogeneous nucleation in the FeNiCrCoCu system but did not report the initial HCP nucleation that we observed [35] while Meesa et al. reported both FCC and HCP structures in comparable amounts with constant cooling in a Fe50Mn30Cr10Co10 system. [36]

Experimental studies in CoCrFeNi revealed that under sufficiently undercooled conditions, a primary metastable BCC phase was observed within which the stable FCC phase nucleated. [37] To the best of our knowledge, this is the first time we report the primary metastable HCP nucleation, which is later superseded by the stable FCC structure.

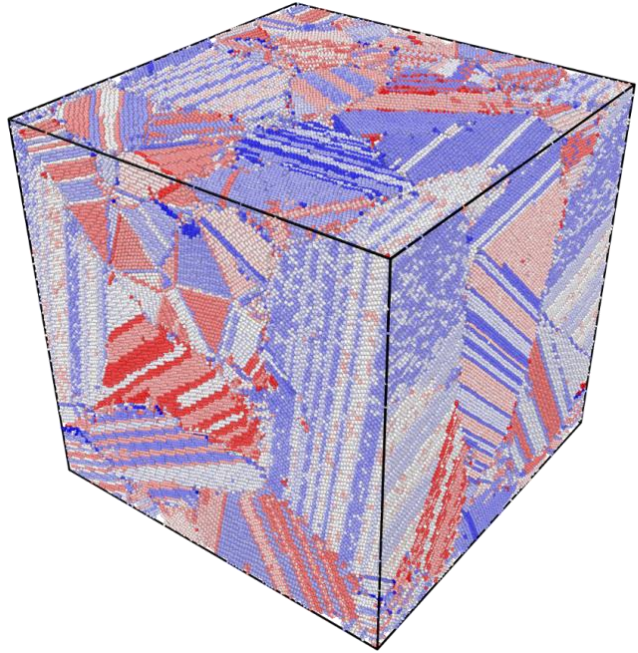

*Figure 1 Grains in the crystal cooled to 1400K (Sim 1400-1)*

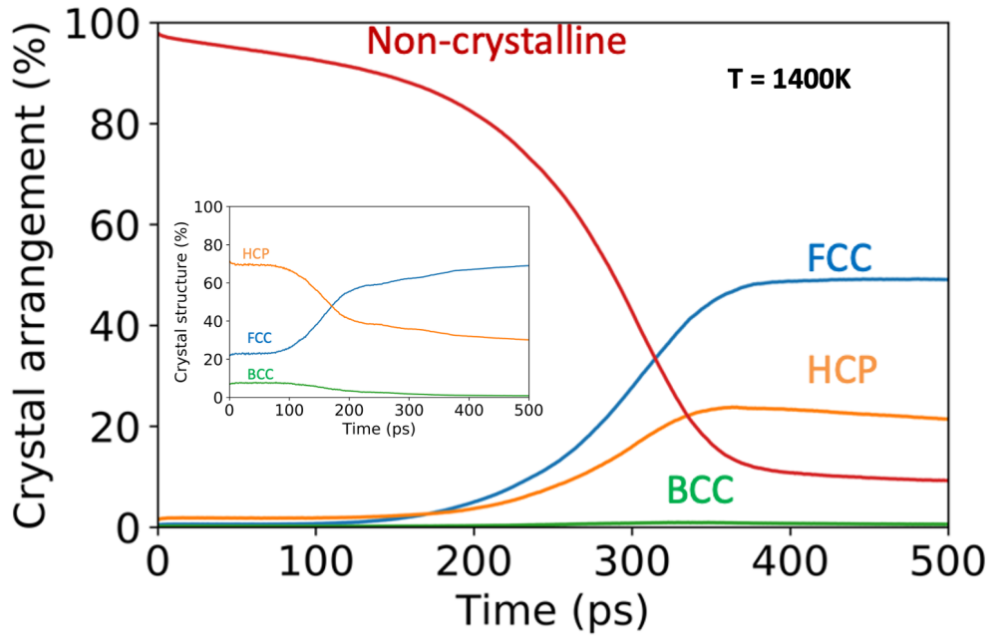

*Figure 2 Amount of non-crystalline, HCP, FCC, and BCC with time. Inset: Amount of HCP, FCC, and BCC as a percentage of the crystallized part at 1400K (Sim 1400-1)*

## Nuclei evolution with time, and classifying nucleation and growth regions

As mentioned above, to track individual nuclei, we performed a cluster analysis over atoms with local crystalline environments, using a cutoff of 3.69Å, this value was chosen because it represents the first minima of the radial distribution function (RDF) of the liquid. Figure 3 shows the number of atoms in different clusters as a function of time for one of the simulations at T=1400 K. We observe the nucleation and dissolution of a large number of nuclei, with a few reaching the critical size and transitioning into the growth regime. These crystals grow and sometimes merge to form large clusters. This analysis shows that at T=1400 K, the critical grain size is approximately 200 atoms. This is consistent with prior work on Au determined by in situ XANES and SAXS experiments. [38] We find similar critical sizes for T=1350 and 1300 K, as shown in the SI.

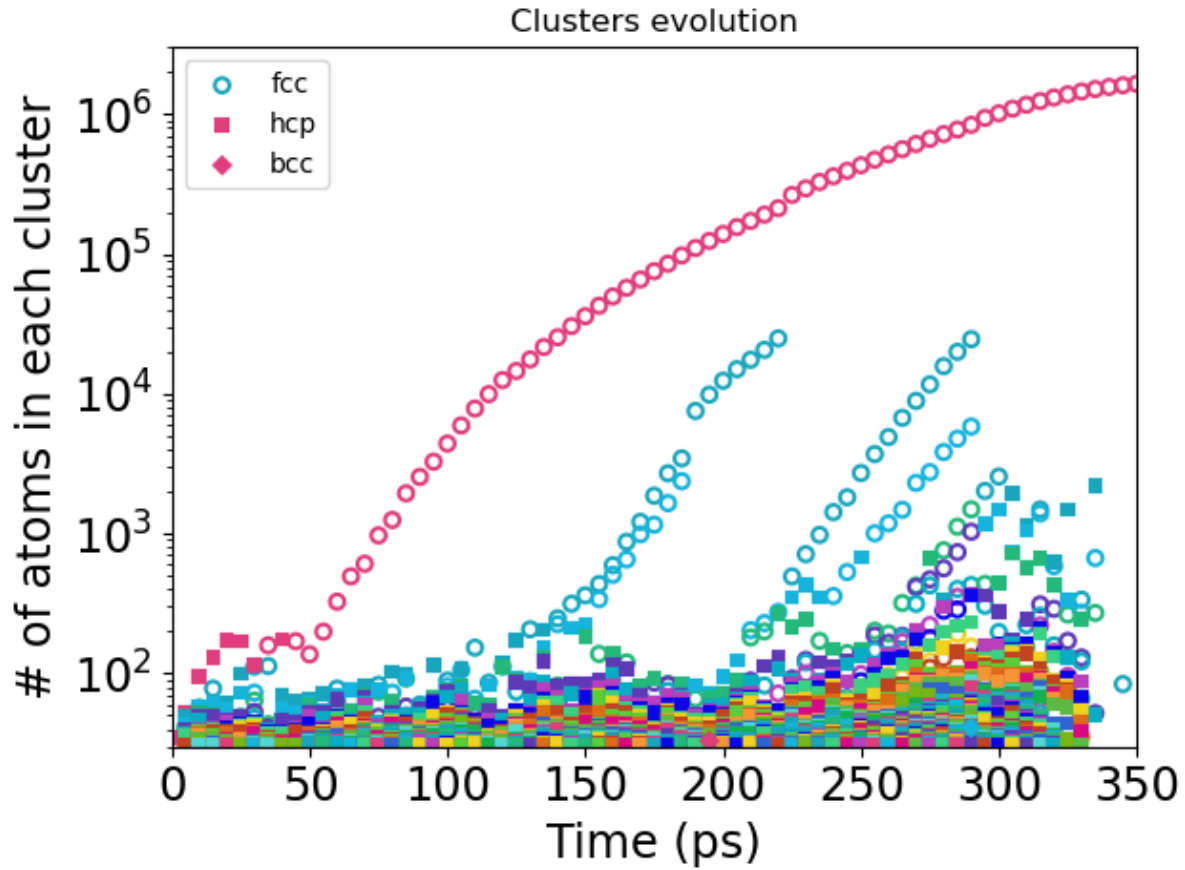

*Figure 3 Number of atoms as a function of time at 1400K (Sim 1400-1). Each color represents a different cluster. Inset: shows the 200-atom cutoff determination Cluster evolution plots for different starting configurations and different temperatures are shown in the Supplementary Information.*

## 4. Composition and energetics during nucleation and growth

### A. Preferential composition during nucleation.

Figure 4 shows the composition of all the sub-critical nuclei identified as a function of their size; each nucleus is represented by five dots of different colors indicating the amount of Co, Cr, Cu, Fe, and Ni. As expected, we observe significant fluctuations in composition with decreasing size. Interestingly, nuclei in the 100-200 atom size tend to be rich in Fe and Co, and Cu-poor.

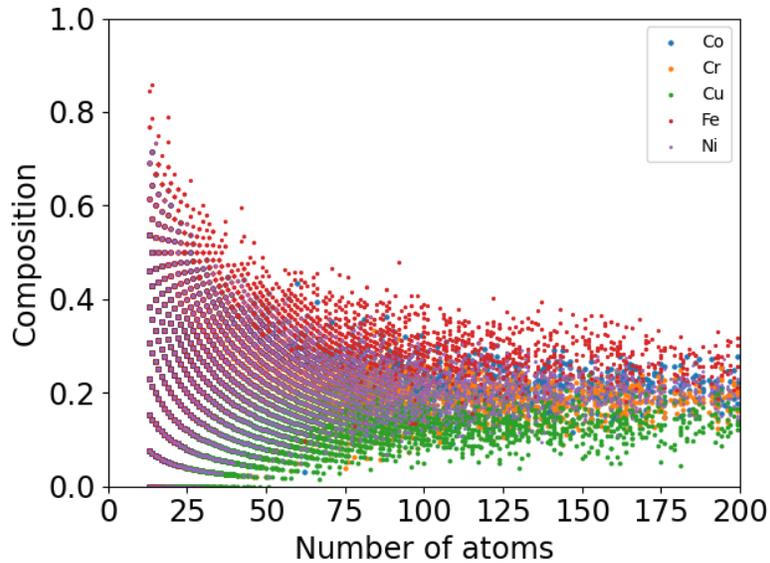

*Figure 4 Composition of all elements at 1400K (Sim 1400-1). Each point represents a nucleus*

To better visualize and understand the dependence of nuclei composition on size and track their time evolution, we used a dimensionality reduction technique (multi-dimensional scaling, MDS, [39-41]) to map the 5-dimensional composition into 2D. Figure 5 shows the composition of each nucleus identified in crystallization simulations at 1400K in three sub-critical size groups (top three rows) and in the growth regime (bottom). Dots are colored by their formation energy on the left panels and by nuclei size on the right. To establish a reference, the gray background shows all possible compositions sampled in steps of 5% (note the spikes in the 2D plot representing compositions high in one of the elements). In addition, to assess random fluctuations due to small sizes, the green background represents compositions obtained by drawing 200-atom systems with equal probabilities assigned to each element. We observe significant deviations from compositionally random nucleation in nuclei ranging from 20 atoms in size all the way to super-critical ones, containing as many as 2000 atoms. Wang et al. studied a CoCrCuFeNi system using MD and also observed Fe-rich nuclei. [18] More importantly, our results are also consistent with experiments on CoCrCuFeNi showing the presence of Cu-rich interdendritic regions during solidification which would lead to Cu-poor dendritic regions that we see in our crystallization study. [42] In our simulation, we severely undercooled the liquid to observe solidification in a reasonable amount of time. We melted the crystal at 3000K, much above the melting temperature ~1800K and cooled it at a rate of 5K/ps to the required temperature and held it at the temperature for 500 ps to observe the solidification process. Wang et al. studied undercooling in CoCrCuFeNi and observed a major high entropy FCC region and a minor metastable Cu-rich region at high undercooling levels during the solidification process. They used EDS to show the higher segregation tendency of Cu and also noted that Ni was more soluble in the Cu-rich phase than Co,Cr, and Fe.[43] All of these observations match well with our results.

The formation energy of the sub-critical nuclei (left panels in Fig. 5) explains their preferential composition. Despite significant fluctuations, we find lower energies (blue) for Fe-rich compositions and higher values closer to high Cu concentrations (red). With increasing nuclei size, the composition fluctuations decrease, and the values converge towards the preferred high-Fe composition. However, the nuclei merge as they grow, and with increasing size, the overall trend is to move towards the center (the overall equiatomic composition). The combination of both these effects is observed in Fig. 5. For larger clusters (marked growth Figure 5(d)-ii) the compositions are a lot closer to the center (equiatomic compositions) and hence we show a zoomed-in plot for growth. The bias for Fe-rich compositions is still observable until about the size of 4000 atoms. We also observe that the smaller clusters (blue) tend to be less equiatomic (off-center) and the larger clusters (red) are more equiatomic (center) showing the growth path that the system takes in the composition space.

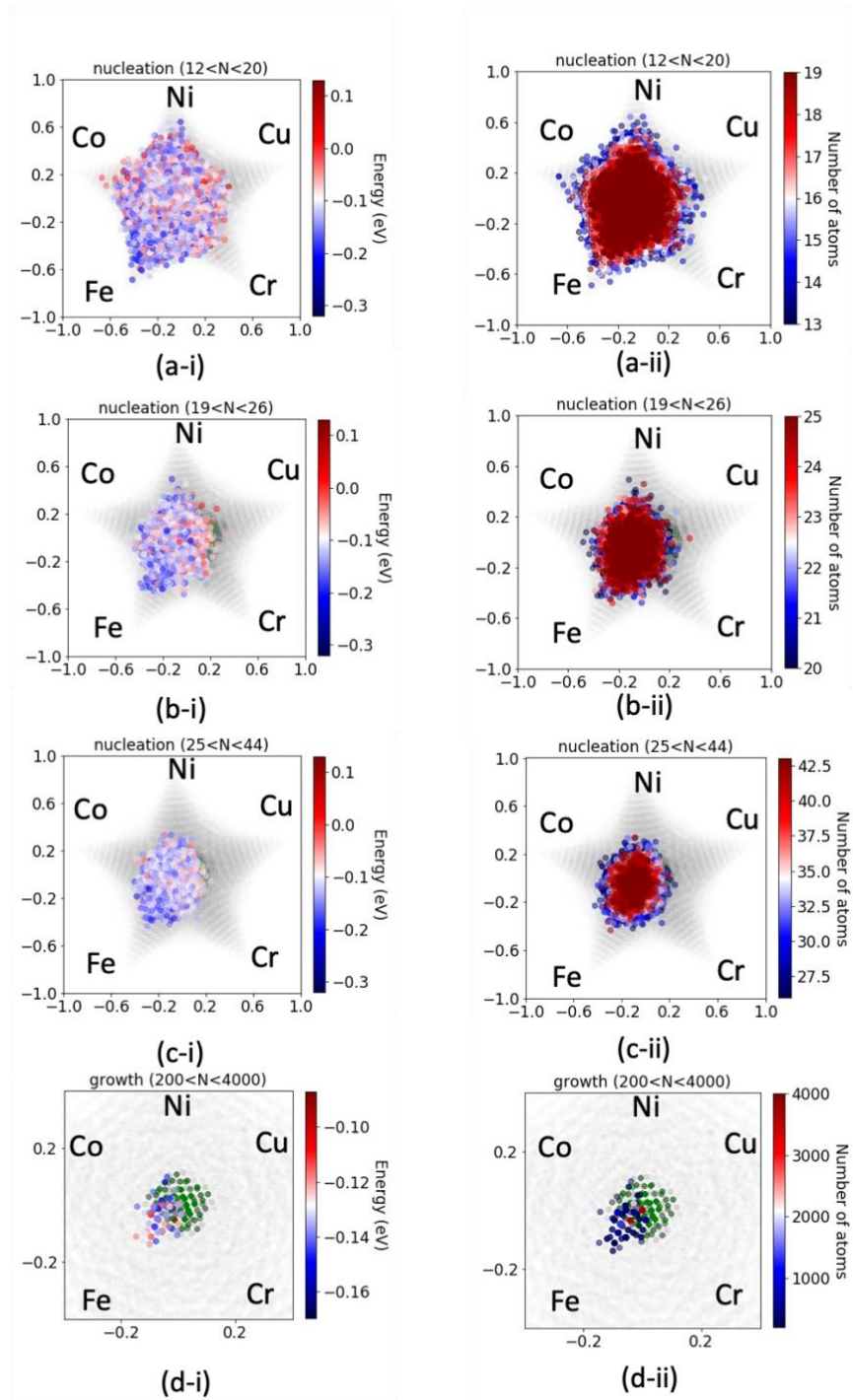

Figure 5 The composition colored by the energy (column 1) and the number (column 2) of atoms at 1400K for different sizes.

## B. Short-range Order (SRO) in CrCoCuFeNi

We now explore whether the compositional bias during nucleation affects the short-range order (SRO) in the crystal. This is important since SRO is known to affect mechanical properties [5]. SRO can be quantified via the Warren-Cowley (WC) parameter that compares the fraction of neighboring atoms of a given type with the probabilities associated with a random structure:

$$a^{nm} = 1 - \frac{P_{nm}}{c_m}$$

$$where\ P_{nm} = \frac{g_{nm}(r < r_0)}{\Sigma_i g_{ni}(r < r_0)}$$

Here $a_{nm}$ is the WC parameter between species n and m, $c_m$ is the composition of the m species and $g_{nm}(r)$ is the radial pair distribution function between the n and m species.

A WC parameter value of 0 represents a perfectly random structure, a negative value shows affinity, and a positive value represents a tendency for the atoms to segregate. Figure 6 shows the average SRO for each pair of elements, at different temperatures. At high temperatures, the equilibrated liquid shows a consistent albeit small SRO. These results indicate the preference of Cr-Fe, Ni-Fe and Fe-Fe bonds over Cr-Cu, Fe-Cu, and Ni-Cu bonds. This correlates with the energy differences discussed in the composition section. The liquid SRO does not change significantly during cooling to the various crystallization temperatures. We observed stabilization of the Co-Cu, Fe-Cu pair and destabilization of Cr-Fe and Fe-Fe pair with cooling from the liquid to the solid. Wang et al. [18] report Fe-Fe preference in the SRO consistent with our observation. We note that they consistently cool their sample at 0.25K/ps while we cool our sample at 5K/ps to the required temperature and maintain the temperature. This does not seem to affect the SRO behavior.

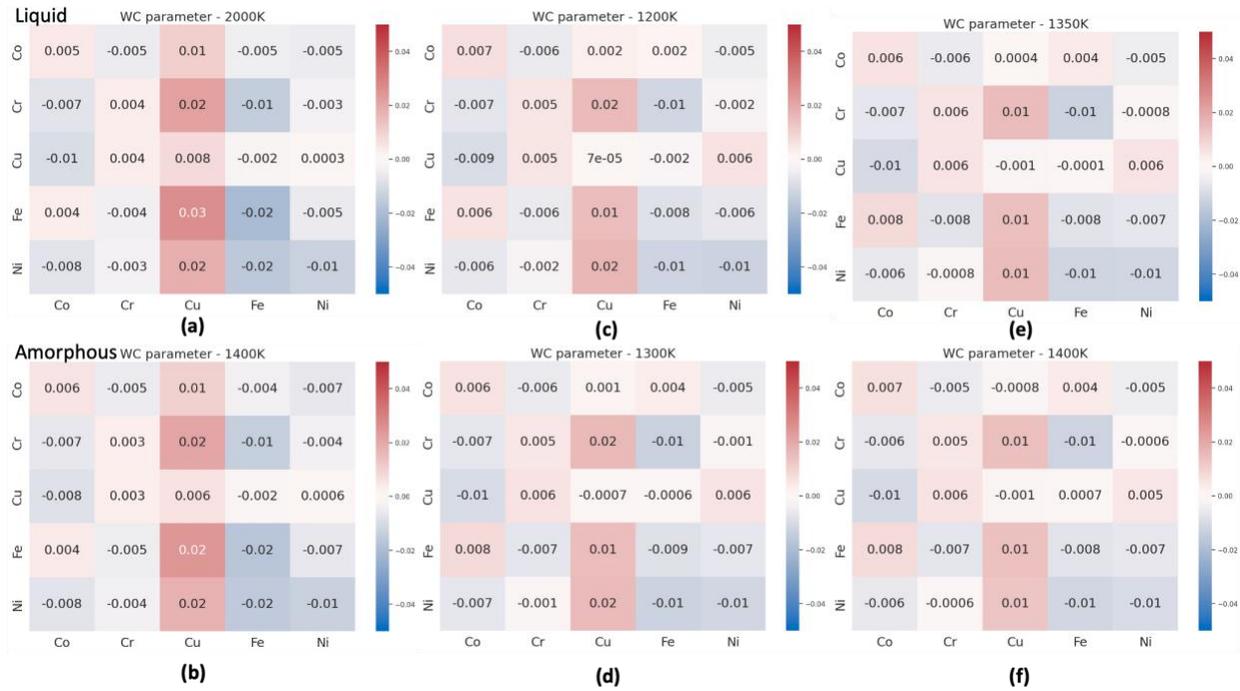

*Figure 6 Warren-Cowley parameters for (a) Liquid at 2000K, (b) Amorphous at 1400K (c) 1200K, (d) 1300K, (e) 1350K, and (f) 1400K.*

## C. Nucleation barriers

To better understand the process of nucleation and growth, Figure 7 shows the energy per atom with respect to the amorphous structure at 1400K (Sim 1400K -1). The overall shape on the internal energy curve is as expected for nucleation and growth processes. Note that these are not free energies (they lack the entropic contribution) which results in negative values. Our results show significant fluctuations in internal energy for small nuclei; these fluctuations decrease with size of the nucleus following the canonical size dependence of $\sqrt{N}$. Interestingly, between sizes of 2,000 and 20,000 atoms, we observe two branches with distinct energies. This phenomenon is observed for a few other starting configurations at 1400K but not for all of them. The chances of multiple branches decrease with temperature. This is expected as at low levels of undercooling, the chances of forming a stable nucleus are lower. This leads to multiple nucleations, few of which grow to give the multiple branches. To better understand this, we have plotted the cluster evolution plots and their corresponding energy vs size plot side by side in the SI. Surprisingly, the high-energy branch prospers and undergoes growth, the underlying reasons for this observation are not clear at this point. This happens when the lower total energy cluster merges with the larger higher total energy cluster.

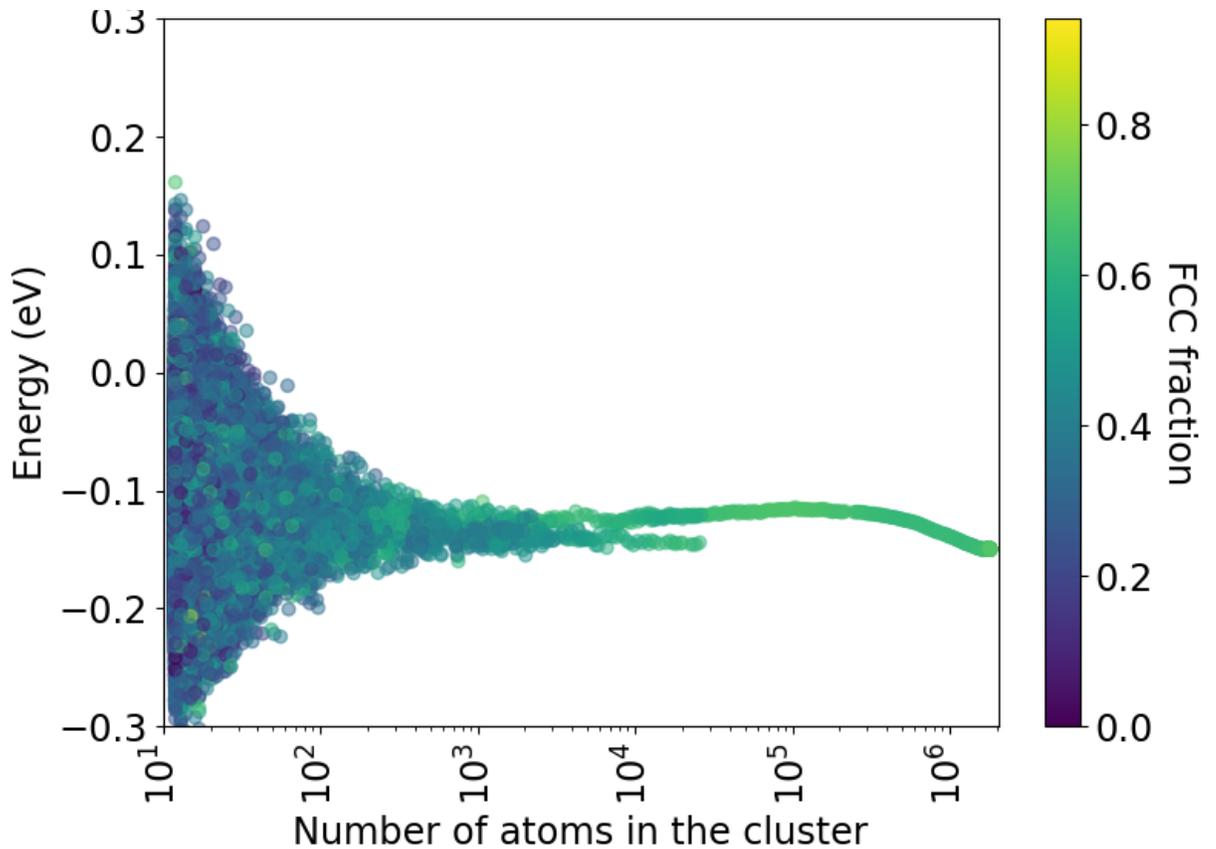

*Figure 7 Average total energy per atom of the clusters as a function of number of atoms colored by the FCC fraction of the cluster at 1400K (Sim 1400K-1)*

## 5. Conclusion

In this study, we explored the nucleation and growth behavior of a model high entropy alloy by cooling the melt and maintaining it below the melting temperature, quantifying the composition bias through MDS analysis and observing deviations from random distribution of compositions. We observed consistent trends in the short-range order plots and the MDS plots, and studied the total energy versus size plots, noting multiple branches for lower levels of undercooling. Interestingy, the lower energy branch did not exhibit growth, a phenomenon we currently cannot explain, while the higher energy branch did. We also found that for our model system, nucleation favors Fe-rich and Cu-poor regions, with this effect decreasing as the size increases, bringing the composition closer to the starting composition. We explained this bias through energetics which makes this study more general and extendable. This work not only provides insights into the behavior of high entropy alloys but also poses new questions and directions for future research in material science.

## 5. Supplementary Material

The supplementary material includes plots for the cluster evolution and the total energy as a function of size for all the different simulations at different temperatures which are referred to in this paper.

**Acknowledgments**

We acknowledge the support from the US National Science Foundation, DMREF program, under Contract Number 1922316-DMR.